\documentclass[12pt]{article}
\usepackage{epsfig,amssymb,amsmath,psfrag,subfigure,rotate}

\allowdisplaybreaks
\newcommand{\insertfig}[2]{\mbox{\epsfxsize=#1cm \epsfbox{#2.pdf}}}

\def \be  {\begin{equation}}
\def \ee  {\end{equation}}
\def \ba  {\begin{eqnarray}}
\def \ea  {\end{eqnarray}}
\def \baa {\begin{eqnarray*}}
\def \eaa {\end{eqnarray*}}
\def \lab #1 {\label{#1}}

\newcommand\re[1]{(\ref{#1})}

\def \matrix #1 {\left(\begin{array}{cc} #1 \end{array}\right)}

\def \tr {\mathop{\rm tr}\nolimits}

\newcommand \vev [1] {\langle{#1}\rangle}
\newcommand \VEV [1] {\left\langle{#1}\right\rangle}
\newcommand \ket [1] {|{#1}\rangle}
\newcommand \bra [1] {\langle {#1}|}

\newcommand{\ft}[2]{{\textstyle\frac{#1}{#2}}}
\begin{document}

\begin{titlepage}

\thispagestyle{empty}

\vspace*{3cm}

\centerline{\large \bf Superpropagator and superconformal invariants}
\vspace*{1cm}

\centerline{\sc A.V. Belitsky$^{a,b}$, S. Caron-Huot$^b$}

\vspace{10mm}

\centerline{\it $^a$Department of Physics, Arizona State University}
\centerline{\it Tempe, AZ 85287-1504, USA}

\vspace{5mm}

\centerline{\it $^b$School of Natural Sciences, Institute for Advanced Study}
\centerline{\it Princeton, NJ 08540, USA}

\vspace{1cm}

\centerline{\bf Abstract}

\vspace{5mm}

We construct a superpropagator in maximally supersymmetric Yang-Mills theory which is invariant off-shell under a chiral half of supersymmetries.
Motivated by the duality with scattering amplitudes in this theory, we apply this superpropagator to supersymmeytric Wilson loop on polygonal
contours. By performing explicit one-loop calculations we confirm the absence of anomalies and verify the duality between the object under study 
and NMHV amplitudes.

\end{titlepage}

\setcounter{footnote} 0

\newpage

\pagestyle{plain}
\setcounter{page} 1

Maximally supersymmetric Yang-Mills theory in the planar limit enjoys hidden symmetries in addition to the explicit superconformal  symmetry of its 
Lagrangian. This property makes this theory quite remarkable as it allows one to use the former as a practical tool to study its physical observables 
at any coupling. One of the discoveries down this route was made recently for the scattering matrix of the model. Namely, it was found that all tree 
amplitudes are invariant under the so-called dual (super)conformal symmetry \cite{DruHenKorSok08,BerMal08}. Quantum mechanically, a number of 
these symmetries gets broken similarly to the space-time (super)conformal symmetries due to infrared divergences of Yang-Mills scattering amplitudes 
generated by the copious emission of massless gauge bosons. The patterns of dual symmetry violation at all order in 't Hooft coupling, while not obvious 
directly for the S-matrix, is well under control making use of another profound revelation which proposes an equivalent description of amplitudes in terms 
of polygonal Wilson loops \cite{AldMal07,KorDruSok07,BranHesTra07}. The contour of the polygon is uniquely determined by the light-like particles' 
momenta involved in scattering process. According to this identification, the bosonic Wilson loop describes the maximal helicity violating (MHV) 
amplitudes. This was demonstrated both at weak coupling to two loop order \cite{DruHenKorSok08a,BerDixKosRoiSprVerVol08} and strong coupling 
\cite{AldMal07,AldMal08}. 

A generalization for amplitudes of arbitrary helicities was suggested by promoting the Wilson loop to chiral superspace by adding Grassmann 
coordinates $\theta^A_\alpha$, that carry and SU(4) index along with the spinor one \cite{MasSki10,CarHuo10}. Thus the superloop $\mathcal{W}$ is 
embedded in a graded space parametrized by the coordinates $\mathcal{Z} = (z_{\alpha\dot\alpha}, \theta^A_\alpha)$. As an object dual to $n$-particle amplitudes, it takes the form of path-ordered product of $n$-segments
\be
\label{SUSYWL}
\mathcal{W}_n = \frac{1}{N_c} \VEV{ P \tr \left( \mathcal{W}_{[1n]} \dots \mathcal{W}_{[32]} \mathcal{W}_{[21]}  \right)  }
\, ,
\ee
where each factor $\mathcal{W}_{j+1j}$ connecting adjacent vertices 
\be
\mathcal{W}_{[i+1i]} = P \exp \left( ig \int^{i+1}_i d \Phi (\mathcal{Z}_i) \right)
\ee
is determined by two superconnections $\mathcal{A}$ and $\mathcal{F}$
\be
\label{dPhi}
d \Phi (\mathcal{Z}_i)
=
\ft12 dz_i^{\dot\alpha \alpha} \mathcal{A}_{\alpha \dot\alpha} (\mathcal{Z}_i) + d \theta^{\alpha A}_i \mathcal{F}_{\alpha A} (\mathcal{Z}_i)
\, .
\ee
These can be constructed systematically\footnote{Recently it was worked out to a rather high order in Grassmann expansion in ref.\ \cite{Gro12}.} and 
are given to the order that suffices for the present consideration by\footnote{Also, in order to simplify notations we use a uniform way of contracting the 
$SL(2)$ indices: undotted indices from upper left to lower right and dotted one lower left to upper right, that is $A^\alpha B_{\alpha\dot\alpha}$ and 
$A_{\dot\alpha} B^{\dot\alpha\alpha}$, and use ket and bra notations $A_\alpha \equiv \ket{A}$, $A^\alpha \equiv \bra{A}$, $A^{\dot\alpha} \equiv |A]$,  
$A_{\dot\alpha} =[A|$. In these notations, contractions of spinors take the conventional form $A^\alpha B_\alpha =\vev{AB}$, $A_{\dot\alpha} 
B^{\dot\alpha} =[AB]$.} \cite{HarShn86}
\begin{align}
\mathcal{A}  
 =& \, 
A +
i \ket{ \theta^A} [ \bar\psi_{A}|
+
\frac{i}{2 !} \ket{\theta^A} \bra{\theta^B} \partial  \bar\phi_{AB} 
-
\frac{1}{3!} \varepsilon_{ABCD} \ket{\theta^A} \bra{\theta^B} D \vev{ \theta^C   \psi^{D}} 
\\[3mm]
& \hfill +
\frac{i}{4!}  \varepsilon_{ABCD} \ket{\theta^A} \bra{\theta^B} D \vev{ \theta^C | F | \theta^{D}}
+
\dots \, ,
\nonumber
\\[3mm]
\mathcal{F}_A =& \, \frac{i}2 \bar\phi_{AB} \ket{\theta^B} 
-
\frac1{3!!}\varepsilon_{ABCD}\ket{\theta^B} \vev{\theta^C \psi^D} 
+
\frac{i}{4!!} \varepsilon_{ABCD}\ket{\theta^B}  \vev{\theta^C | F |  \theta^D}+\ldots
\, .
\end{align}
By construction the transformation under the chiral Poincar\'e supersymmetry takes the following form~\cite{BelKorSok11}
\begin{align}
\label{SUSYtransformAandF}
\delta_\varepsilon \mathcal{A}_{\alpha\dot\alpha} & = \partial_{\alpha\dot\alpha} \omega 
+ i g [\omega,  \mathcal{A}_{\alpha\dot\alpha} ]
+ \Omega_{\alpha\dot\alpha}\,,\qquad
\\[2mm]\notag
\delta_\varepsilon {\mathcal F}_{\alpha A} & = \partial_{\alpha A}\omega + i g [\omega,  \mathcal{F}_{\alpha A} ]
\, .
\end{align}
Here $\omega$ is the field-dependent gauge transformation parameter
\be
\label{Gaugeomega}
\omega 
=
\vev{\varepsilon^{A} \theta^B}\left[
- \frac{i}{2!} \bar\phi_{AB}
+
\frac{1}{3!} \varepsilon_{ABCD} \vev{\theta^C \psi^D}
-
\frac{i}{4!} 
\varepsilon_{ABCD} \vev{\theta^C | F | \theta^D }
+
\dots
\right]
\,
\ee
while the last term in the variation of $\mathcal{A}$ stands for the field equation of motion whose presence is a consequence of the fact that the 
$\mathcal{N}=4$ supersymmetric algebra closes only on-shell in the space spanned by the quantum fields. Namely,
\ba
\label{EOMOmega}
\Omega_{\alpha\dot\alpha}
=
-
2
\varepsilon_{ABCD}
(\varepsilon^{\beta A} \theta_\beta^B) \theta^C_\alpha
\left[
\frac{1}{3!} (\Omega_{\rm f})_{\dot\alpha}^D
-
\frac{i}{4!} \theta_\gamma^D (\Omega_{\rm g})_{\dot\alpha}{}^\gamma
+ \dots
\right]
\, 
\ea
written in terms of gaugino and gauge field equations of motion
\be
\label{EOMoperators}
(\Omega_{\rm f})_{\dot\alpha}^A 
= D_{\dot\alpha}{}^\gamma \psi^A_\gamma
\, , \qquad
(\Omega_{\rm g})_{\dot\alpha}{}^\alpha
= D_{\dot\alpha}{}^\gamma F_\gamma{}^\alpha
\, .
\ee

The above transformation law ensures supersymmetry of the Wilson loop at the classical level.  However, a difficulty arises at the quantum level.
As was demonstrated in ref.\  \cite{BelKorSok11}, the presence of light-cone singularities in Feynman graphs requires introduction of a regularization 
procedure to make them well-defined at intermediate steps.  Making use of the Four Dimensional Helicity scheme \cite{BerFreDixWon02} that suits well 
computation of scattering amplitudes in spinor-helicity formalism and which is also used on the Wilson loop side, one finds that the superloop becomes 
anomalous. This was understood as a consequence of the insertion of the aforementioned equations of motion \re{EOMoperators} into the loop contour 
which induce a finite contribution upon the cancellation of order $\varepsilon$ effects by the light-cone poles in $1/\varepsilon$ triggered by the loop 
integration \cite{BelKorSok11,Bel12}. Note that this problem would be absent if the variation involved the equations of motion of the $D$-dimensional 
theory one is working in, the issue here being that with the known operator only the 4-dimensional part of the equations of motion gets generated as 
explained in \cite{BelKorSok11}.

Needless to say, such a violation of supersymmetry would be fatal for the duality. Recall that on the amplitude side of the equivalence, the dual
Poincar\'e supersymmetry is trivially preserved being merely a way the dual variables were introduced in a supertranslationally invariant manner.

In the absence of a known supersymmetric regularization of the operator \re{SUSYWL}, a standard fallback strategy is to construct finite symmetry-restoring 
counterterms order-by-order in perturbation theory.  An analogous well-studied example is the case of conformal operators in a conformal theory, where 
conformal symmetry is violated by the use of a dimensional regulator but can be restored by a suitable finite renormalization. Using insight from the conformal 
symmetry, a finite renormalization of the superloop was thus constructed in \cite{Bel12} and argued to restore the duality, to this order. The absence of 
anomalies can also be demonstrated using the so-called framing regulator following \cite{BeiHeSchVer12}, which reproduces the NMHV tree amplitudes.

In this work we follow a different strategy and analyze whether the perturbative expression for the supersymmetric Wilson loop can be rewritten in an explicitly super-Poincar\'e  invariant form. Presently we will perform a feasibility analysis at one loop order and thus limit ourselves to the Grassmann degree-four terms which
are dual to tree NMHV amplitudes. Then, the only type of Feynman graphs that can contribute involve a propagator connecting different sites of the 
superloop. Thus the main object of our consideration will be the propagator for the superconnection $d \Phi$ introduced
in Eq.\ \re{dPhi},
\be
\mathcal{G}_{12} = \vev{d \Phi (\mathcal{Z}_1) d \Phi (\mathcal{Z}_2)}
= \int [DX] d \Phi (\mathcal{Z}_1) d \Phi (\mathcal{Z}_2) {\rm e}^{i S}
\, ,
\ee
where the coordinates $\mathcal{Z}_i$ determine the attachment of the propagator to the loop segments. This two-point correlation function is defined as
path integral with the weight given by the total gauge-fixed action $S = S_{\rm cl} + S_{\rm gf} + S_{\rm gh}$. An explicit expression for $\mathcal{G}_{12}$ 
can be constructed in terms of superfields' components and reads
\ba
\label{OrigProp}
\mathcal{G}_{12} 
=
\bigg[
\!\!\!&-&\!\!\! \ft{1}{24} \bra{\theta_2} dz_1 \partial_1 \ket{\theta_2} \left[  \bra{\theta_2} dz_2 \partial_2 \ket{\theta_2} - 6 \vev{d \theta_2 \theta_2} \right]
+ \ft{1}{6} \bra{\theta_1} dz_1 \partial_1 \ket{\theta_2} \left[ \bra{\theta_2} dz_2 \partial_2 \ket{\theta_2} - 4 \vev{d \theta_2 \theta_2} \right]
\nonumber\\
&-&\!\!\!
\ft{1}{8}
\left[ \bra{\theta_1} dz_1 \partial_1 \ket{\theta_1} - 2 \vev{d \theta_1 \theta_1} \right]
\left[ \bra{\theta_2} dz_2 \partial_2 \ket{\theta_2} - 2 \vev{d \theta_2 \theta_2} \right]
\bigg]
\frac{\Gamma (1 - \varepsilon)}{16 \pi^{2 - \varepsilon} [- z_{12}^2]^{1 - \varepsilon}}
\, ,
\ea
where we implicitly adopted the Feynman gauge. Here and below we will keep all expressions in $D$-dimensions so that we can clearly identify the
differences from the earlier analyses. It will turn out that the final result will have a very soft light-cone behavior and does not require any regularization
at all. Let us see how this object transforms under chiral supersymmetry. Making use of the transformation laws \re{SUSYtransformAandF}, one immediately 
finds that
\ba
\delta_\varepsilon \mathcal{G}_{12} 
&=& 
{\rm d}_2 
\vev{
d \Phi (\mathcal{Z}_1) \omega (\mathcal{Z}_2)}
+
\ft14 dz_1^{\dot\alpha\alpha}  dz_2^{\dot\beta\beta}  
\vev{\mathcal{A}_{\alpha\dot\alpha} (\mathcal{Z}_1) \Omega_{\beta\dot\beta} (\mathcal{Z}_2)}
+ (1 \leftrightarrow 2)
\nonumber\\
&+& \vev{d \Phi (\mathcal{Z}_1) d \Phi (\mathcal{Z}_2) i (\delta_\varepsilon S_{\rm gf})}
\, ,
\ea
where ${\rm d}_i$ is an external super-differential
\be
{\rm d}_i = \ft12 dz_i^{\dot\alpha \alpha} \partial_{i \alpha\dot\alpha} + d \theta_i^{\alpha A} \partial_{i \alpha A}
\, ,
\ee
and we accounted for the fact that for the tree propagator one can omit the ghost portion of the Lagrangian, while the classical part is obviously
being supersymmetric invariant on its own. We are now in a position to calculate each term in the above equation. First, using the explicit component 
form for the superconnections and $\omega$, we find the following expression for the first line
\ba
\vev{\mathcal{A}_{\alpha\dot\alpha} (\mathcal{Z}_1) \omega (\mathcal{Z}_2)}
&=&
\frac{\Gamma (2 - \varepsilon)}{4 \pi^{2 - \varepsilon}} 
\frac{\varepsilon_{ABCD}}{[- z_{12}^2]^{2 - \varepsilon}}
\vev{\varepsilon^C \theta_2^D}
\left[
\ft{1}{12} \theta^A_{2 \alpha} \theta^{B, \beta}_2 - \ft13 \theta^A_{1 \alpha} \theta^{B, \beta}_2 + \ft12 \theta^A_{1 \alpha} \theta^{B \beta}_1
\right] 
(z_{12})_{\dot\alpha\beta}
\, , \nonumber\\
\vev{\mathcal{F}_{\alpha A} (\mathcal{Z}_1) \omega (\mathcal{Z}_2)}
&=&
\frac{\Gamma (1 - \varepsilon)}{16 \pi^{2 - \varepsilon}} 
\frac{\varepsilon_{ABCD}}{[- z_{12}^2]^{1 - \varepsilon}}
\vev{\varepsilon^C \theta_2^D} \theta_{1 \alpha}^B
\, .
\ea
Next, we make use of the following results for regularized propagators involving equations of motion
\ba
\label{epsilonEOM}
\vev{\bar\psi_{\dot\alpha A} (z_1) (\Omega_{\rm f})_{\dot\beta}^B (z_2)}
&=& 
i \delta_A^B \varepsilon \frac{\Gamma (2 - \varepsilon)}{\pi^{2 - \varepsilon}} \frac{\varepsilon_{\dot\alpha\dot\beta}}{[- z_{12}^2]^{2 - \varepsilon}}
\, , \nonumber\\
\vev{A_{\alpha\dot\alpha} (z_1) (\Omega_{\rm g})_{\dot\beta\beta} (z_2)}
&=&
\frac{i \Gamma (2 - \varepsilon)}{4 \pi^{2 - \varepsilon}} 
\left\{
\partial_{1 \alpha\dot\alpha} \frac{z_{12}{}_{\beta\dot\beta}}{[- z_{12}^2]^{2 - \varepsilon}} 
+ 
4 \varepsilon \frac{\varepsilon_{\dot\alpha \dot\beta} \varepsilon_{\alpha\beta}}{[- z_{12}^2]^{2 - \varepsilon}}
\right\}
.
\ea
The swindle will consist in ignoring $O (\varepsilon)$ effects. The latter when kept in the Wilson loop induce finite contibutions being compensated by the 
light-cone divergences emerging from integrations in the vicinity of the cusps. In other words, if one deals with a non-regularized integrand, one would not 
even notice these extra terms and thus we deduce
\ba
\ft14 dz_1^{\dot\alpha\alpha}  dz_2^{\dot\beta\beta}  
\vev{\mathcal{A}_{\alpha\dot\alpha} (\mathcal{Z}_1) \Omega_{\beta\dot\beta} (\mathcal{Z}_2)}
&=&
- \frac{1}{12} 
{\rm d}_1
\vev{\varepsilon \theta_2}
\bra{\theta_2} dz_2 \partial_2 \ket{\theta_2}
\frac{\Gamma (1 - \varepsilon)}{16 \pi^{2 - \varepsilon}} \frac{1}{[- z_{12}^2]^{2 - \varepsilon}}
\, ,
\ea
where $\bra{\theta_2} dz_2 \partial_2 \ket{\theta_2} = \theta_2^\alpha (dz_2)_{\alpha\dot\alpha} (\partial_2)^{\dot\alpha\beta} 
\theta_{2 \beta}$ following the conventions we have adopted earlier.

\begin{figure}[t]
\begin{center}
\mbox{
\begin{picture}(0,120)(60,0)
\put(0,0){\insertfig{5}{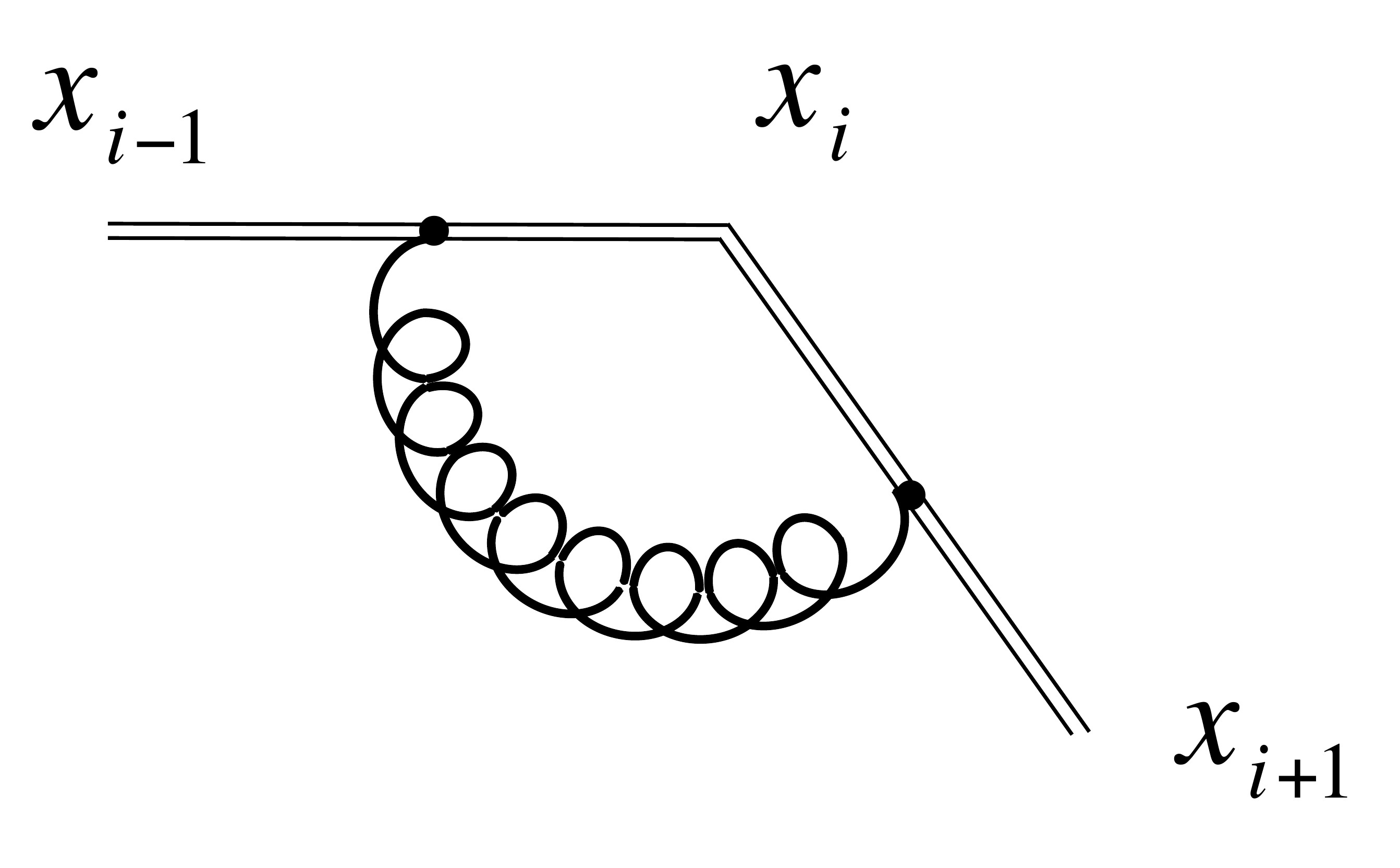}}
\end{picture}
}
\end{center}
\caption{ \label{Vertex} Exchange correction to the single cusp.}
\end{figure}

Last but not least, since the propagator is not gauge invariant by itself, we have to incorporate the contribution from the variation of the action.
Since the calculation is done in the Feynman gauge, the variation of the gauge-fixing terms (with $\xi = 1$)
\be
i \delta_\varepsilon S_{\rm  gf} 
= 
\frac{1}{4 \xi} \int d^{4 - 2 \varepsilon} z (\partial^{\dot\alpha\alpha} A_{\alpha\dot\alpha} (z)) (\varepsilon^{\beta B} \partial_{\beta\dot\beta} \bar\psi^{\dot\beta}_B)
\, ,
\ee
immediately yields yet another contribution 
\be
\vev{d \Phi (\mathcal{Z}_1) d \Phi (\mathcal{Z}_2) i (\delta_\varepsilon S_{\rm gf})}
=
- \frac{2}{3 \xi} {\rm d}_1 \vev{\varepsilon \theta_2} \vev{d \theta_2 \theta_2} \frac{\Gamma (1- \varepsilon)}{16 \pi^{2 - \varepsilon}} \frac{1}{[- z_{12}^2]^{1 - \varepsilon}}
+ (1 \leftrightarrow 2)
\, .
\ee
Summing everything together, we find that the chiral variation is given by a total differential 
\be
\label{PropSuperVariation}
\delta_\varepsilon \mathcal{G}_{12} = {\rm d}_2 \mathcal{X}_{12} + {\rm d}_1\mathcal{X}_{21}
\, ,
\ee
of a degree-four Grassmann valued function $\mathcal{X}_{12}$ of a very special form
\ba
\mathcal{X}_{12} 
\!\!\!&=&\!\!\! 
\vev{\varepsilon \theta_2}
\left[
\vev{d \theta_1 \theta_1} - \frac{1}{12} \bra{\theta_2} dz_1 \partial_1 \ket{\theta_2}
+  \frac{1}{3} \bra{\theta_1} dz_1 \partial_1 \ket{\theta_2}  - \frac{1}{2} \bra{\theta_1} dz_1 \partial_1 \ket{\theta_1}
\right] \frac{\Gamma (1 - \varepsilon)}{16 \pi^{2 - \varepsilon}} \frac{1}{[- z_{12}^2]^{1 - \varepsilon}}
\nonumber\\
&+&\!\!\! 
\vev{\varepsilon \theta_1}
\left[
- \frac{2}{3} \vev{d \theta_1 \theta_1} + \frac{1}{4} \bra{\theta_1} dz_1 \partial_1 \ket{\theta_1}
\right] \frac{\Gamma (1 - \varepsilon)}{16 \pi^{2 - \varepsilon}} \frac{1}{[- z_{12}^2]^{1 - \varepsilon}}
\, .
\ea
Since the one-loop Wilson loop is given by a sum over all segments, the total derivatives can be safely gauged away. Making use of this result, 
we can define a new propagator $\widetilde{\mathcal{G}}_{12}$ whose chiral variation vanishes exactly, i.e., $\delta_\varepsilon \widetilde{\mathcal{G}}_{12}
= 0$. This requires introduction of additive terms to the original propagator $\mathcal{G}_{12}$ such that their chiral variation coincides with minus the 
right-hand side of Eq.\ \re{PropSuperVariation}. A computation then yields
\ba
\label{SUSprop}
\widetilde{\mathcal{G}}_{12}
=
\!\!\!&-&\!\!\! 
\ft{1}{2}
\left[
\bra{\theta_{12}} dz_1 z_{12} \ket{\theta_{12}} \vev{d \theta_2 \theta_{12}}
+
\vev{d \theta_1 \theta_{12}} \bra{\theta_{12}} dz_2 z_{12} \ket{\theta_{12}} 
\right]
\frac{\Gamma (2 - \varepsilon)}{16 \pi^{2 - \varepsilon} [- z_{12}^2]^{2 - \varepsilon}}
\\
&+&\!\!\!
\ft{2}{3} \vev{d \theta_1 \theta_{12}} \vev{d \theta_2 \theta_{12}} \frac{\Gamma (1 - \varepsilon)}{16 \pi^{2 - \varepsilon} [- z_{12}^2]^{1 - \varepsilon}}
+
\ft{1}{6} \bra{\theta_{12}} dz_1 z_{12} \ket{\theta_{12}} \bra{\theta_{12}} dz_2 z_{12} \ket{\theta_{12}}
\frac{\Gamma (3 - \varepsilon)}{16 \pi^{2 - \varepsilon} [- z_{12}^2]^{3 - \varepsilon}}
\, . \nonumber
\ea
This manifestly supersymmetric propagator is the main result of this paper.  It differs from the original one \re{OrigProp} by a super-gauge transformation 
accompanied by the additive order-$\varepsilon$ terms in (\ref{epsilonEOM}).

\begin{figure}[t]
\begin{center}
\mbox{
\begin{picture}(0,150)(170,0)
\put(0,10){\insertfig{12}{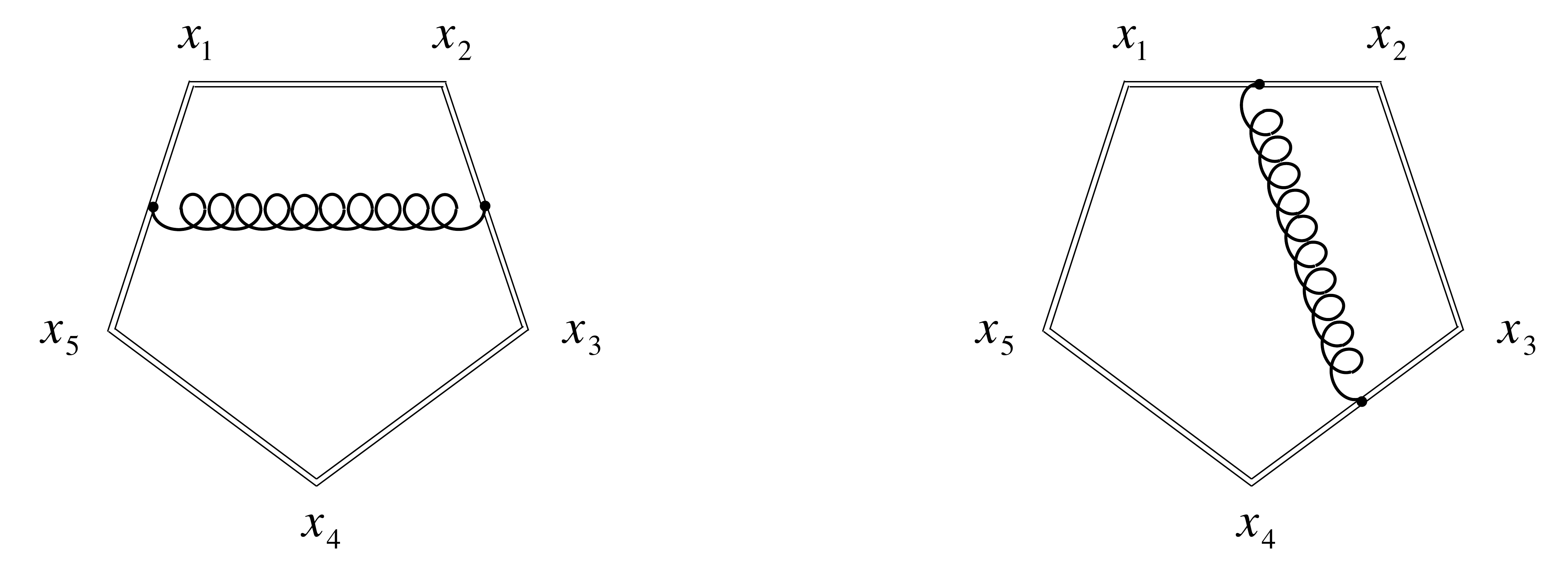}}
\put(60,-10){(a)}
\put(265,-10){(b)}
\end{picture}
}
\end{center}
\caption{ \label{Pentagon} Nonvanishing one-loop graphs for the $\chi_1^4$ component of the pentagon.}
\end{figure}

In what follows, let us perform a few perturbative tests making use of this propagator. We focus on the notorious $\chi_1^4$-component which 
received anomalous contribution from exchange graphs involving nearest and next-to-nearest links \cite{BelKorSok11}. The consideration for 
other components in the Grassmann expansion is completely analogous and we will present a formula summarizing them in Eq.\ \re{SuperConfInv}.

Making use of Eq.\ \re{SUSprop}, one finds immediately that dressing up the cusp with the propagator as shown in Fig.\ \ref{Vertex}, one finds that
each term in the expression \re{SUSprop} individually generates finite contributions, i.e., $\int_0^1 dt \, t^\varepsilon = 1+ O(\varepsilon)$, as compared to 
ultraviolet divergent integrals made formally finite upon the use of a dimensional regulator for the original component propagators in Eq.\ \re{OrigProp}, 
$\int_0^1 dt \, t^{-2 + \varepsilon} = -1 + O(\varepsilon)$. Moreover, the entire former contribution vanishes (after setting $\varepsilon = 0$) as a consequence 
of the cancellation between the four terms in Eq.\ \re{SUSprop}, $\ft13 - \ft12 - \ft12 + \ft23 = 0$. Having observed that, the next question is whether 
the previously non-vanishing and purely anomalous four-point Wilson loop receives any finite contributions at this order. An inspections immediately 
shows that exchange graphs between the next-to-nearest links vanish individually for the same token as observed for the vertex graphs, thus the NMHV 
four-point amplitude is zero as it is supposed to be.

Let us now turn to less trivial cases and start with the pentagon. The only novanishing diagrams contributing to its $\chi_1^4$ component are displayed 
in Fig.\ \ref{Pentagon}. Their calculation gives for (a) and (b), respectively,
\begin{align}
\mathcal{W}^{(\rm a)}_5 
&= 
- \frac{a \chi_1^4}{48} \frac{\vev{52}^2}{\vev{51}^2 \vev{12}^2} \frac{x_{35}^2}{x_{52}^2 x_{13}^2}
\, , \\
\mathcal{W}^{(\rm b)}_5 
&=
\frac{a \chi_1^4}{48}  \frac{\vev{52} \vev{23} \bra{5} x_{53} |3]}{\vev{51}^2 \vev{12}^2} \frac{\log (x_{14}^2/x_{24}^2)}{(x_{14}^2 - x_{24}^2)^2}
\nonumber\\ 
&-
\frac{a \chi_1^4}{48} \frac{\bra{5} x_{53} |3]}{\vev{51}^3 \vev{12}^2 x_{13}^2 (x_{14}^2 - x_{24}^2)}
\left\{
\vev{52}^2 \vev{13} + \vev{23} \bra{5} x_{53} |3] \left[ \frac{\vev{53} \vev{12}}{x_{14}^2} + \frac{\vev{23} \vev{51}}{x_{24}^2} \right]
\right\}
\, , \nonumber
\end{align}
such that adding the exchange diagram (a) to the sum of (b) and its mirror image (with the logarithms canceling between the latter two), we get the $\chi_1^4$ 
component of the superconformal $R$-invariant (see Eq.\ \re{Rsuperconf} below). Again, it is anomaly-free.

For more solid confirmation of the above observations, we computed the hexagon as well. The results are summarized in the following 
equations,
\begin{align}
\vev{\mathcal{W}_{6;1}^{({\rm a})}}
&=
- \frac{a \chi_1^4}{48} \frac{\vev{62}^2}{\vev{61}^2 \vev{12}^2}
\frac{x_{63}^2}{x_{13}^2 x_{62}^2}
\, ,\\[3mm]
\vev{\mathcal{W}_{6;1}^{({\rm b})}}
&=
\frac{a \chi_1^4}{48}  \frac{\vev{62} \vev{23} \bra{6} x_{63} |3]}{\vev{61}^2 \vev{12}^2} \frac{\log (x_{14}^2/x_{24}^2)}{(x_{14}^2 - x_{24}^2)^2}
\nonumber\\
&-
\frac{a \chi_1^4}{48} \frac{1}{\vev{61}^3 \vev{12}^2}
\frac{\bra{6} x_{63} |3]}{x_{13}^2 (x_{14}^2 - x_{24}^2)}
\bigg\{
\vev{13} \vev{62}^2
+
\vev{23} \bra{6} x_{63} |3]
\left[
\frac{\vev{12} \vev{63}}{x_{14}^2} + \frac{\vev{23} \vev{61}}{x_{24}^2}
\right]
\bigg\}
\, , \nonumber\\[3mm]
\vev{\mathcal{W}_{6;1}^{({\rm c})}}
&=
- 
\frac{a \chi_1^4}{48}  \frac{\vev{62} \vev{23} \bra{6} x_{63} |3]}{\vev{61}^2 \vev{12}^2} \frac{\log (x_{14}^2/x_{24}^2)}{(x_{14}^2 - x_{24}^2)^2}
+
 \frac{a \chi_1^4}{48}  \frac{\vev{62} \vev{56} \bra{2} x_{52} |5]}{\vev{61}^2 \vev{12}^2} \frac{\log (x_{15}^2/x_{25}^2)}{(x_{15}^2 - x_{25}^2)^2}
\nonumber\\
&+ 
\frac{a \chi_1^4}{48}  \frac{1}{\vev{61}^3 \vev{12}^3}
\bigg\{
\vev{14} \vev{62}^2 \frac{\vev{23} \vev{61} [34] - \vev{12} \vev{56} [45]}{(x_{14}^2 - x_{24}^2) (x_{52}^2 - x_{51}^2)}
\\
&\qquad\qquad\qquad\quad
-
\frac{\bra{1} x_{14} |4]}{x_{14}^2 x_{52}^2 - x_{24}^2 x_{51}^2}
\bigg[
\frac{\vev{23}^2 \bra{6} x_{63} |3]^2}{x_{14}^2 - x_{25}^2}
\left(
\frac{\vev{12} \vev{64}}{x_{14}^2} + \frac{\vev{61} \vev{24}}{x_{24}^2}
\right)
\nonumber\\
&\qquad\qquad\qquad\qquad\qquad\qquad\qquad\ 
+
\frac{\vev{56}^2 \bra{2} x_{52} |5]^2}{x_{52}^2 - x_{51}^2}
\left(
\frac{\vev{12} \vev{64}}{x_{52}^2} + \frac{\vev{61} \vev{24}}{x_{51}^2}
\right)
\bigg]
\bigg\}
\, . \nonumber
\end{align}
Summing up these together yields a well-known expression for the $\chi_1^4$ component of the tree NMHV amplitude.

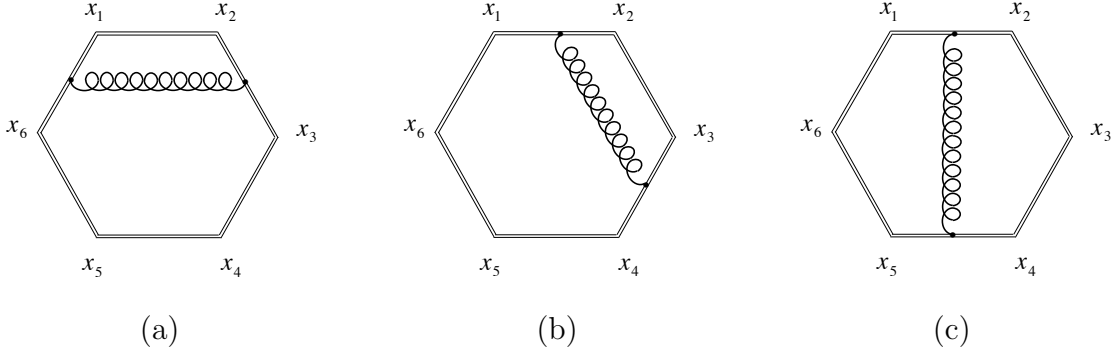
\begin{figure}[t]
\begin{center}
\mbox{
\begin{picture}(0,150)(215,0)
\put(0,20){\insertfig{15}{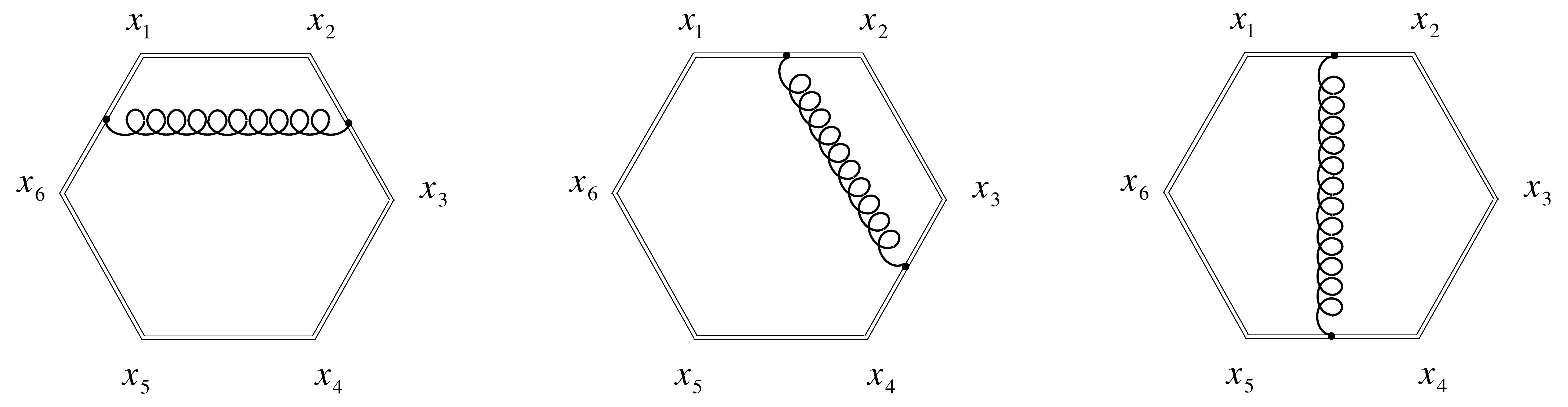}}
\put(55,0){(a)}
\put(205,0){(b)}
\put(355,0){(c)}
\end{picture}
}
\end{center}
\caption{ \label{Hexagon} Exchange graphs contributing to the $\chi_1^4$ structure of the haxagon.}
\end{figure}

We can immediately generalize these considerations to any number of cusps. Namely, it is based on the observation that while all exchange 
Feynman diagrams connecting all but nearest-neighbor segments produce the same result when computed either for the old or new propagator 
for the $\chi_1^4$ Grassmann structure, the difference between the $[n1]$-$[23]$ exchange diagram computed with the supergauge-transformed 
propagator \re{SUSprop} and the vertex correction to cusps in points $\mathcal{Z}_1$ and $\mathcal{Z}_2$ evaluated with the old one \re{OrigProp} 
are given by
\be
\Delta \vev{\mathcal{W}_{n;1}}
=  
\frac{a}{48} \chi_1^4
\frac{[n2]}{[n1][12]} \frac{\vev{n2}^3}{\vev{n1}^3 \vev{12}^3}
\, .
\ee
This exactly cancels the conformal anomaly \cite{Bel12} that broke the duality between the super Wilson loop and NMHV amplitudes. The rest of the 
exchange graphs are conformally invariant and their sum yields the $\chi_1^4$ component of the $R$-invariants defining the tree NMHV superamplitude, 
i.e., $\sum_{1 < q < r < n} R_{n;qr} |_{\chi_1^4}$. Thus generalizing this finding to the complete one-loop super Wilson loop, we conclude that
\be
\label{SuperConfInv}
\ft12 \sum_{i \neq j} \int_0^1 dt_i \int_0^1 dt_j \, \widetilde{\mathcal{G}}_{ij} (x_{[ii+1]} (t_i), x_{[jj+1]} (t_j)) = \frac{1}{96 \pi^2} \sum_{1<q<r<n} R_{n;qr}
\, ,
\ee
(with $x_{ii+1} (t_i) = x_i - t_i x_{ii+1}$) where the superconformal invariant is defined by the equation \cite{DruHenKorSok08}
\be
\label{Rsuperconf}
R_{n;qr} 
= 
\frac{\delta^4 (\vev{n, q-1, q, r-1} \chi_r + \ {\rm cyclic})}{\vev{q-1, q, r-1, r} \vev{q, r-1, r, n} \vev{r-1, r, n, q-1} \vev{r, n, q-1, q} \vev{n, q-1, q, r-1}}
\, ,
\ee
written in terms of momentum twistors $Z^a_j = (\lambda^\alpha_j, x_j^{\dot\alpha\alpha} \lambda_{j \alpha})$, with angle-brackets being
$\vev{ijkl} = \varepsilon_{abcd} Z^a_i Z^b_j Z^c_k Z^d_l$.
This is the space-time analogue of the twistor-space result in Ref.\ \cite{MasSki10}.

For components other than $\chi_1^4$ it can be verified that the cusp diagrams as in Fig.\ \ref{Vertex} continue to vanish exactly, and the other diagrams
are convergent and thus that the Wilson loop is manifestly anomaly free. Based on the general recursion arguments presented in ref.~\cite{CarHuo10},
which assumed only supersymmetry together with the conjectured cancellation of non-rational terms (logarithms) which was repeatedly confirmed above,
it is thus virtually certain that the complete NMHV tree amplitude will be reproduced correctly.

We have shown in the present note that one can construct an explicitly supersymmetric form for the tree superpropagator of superconnections
defining the one-loop super Wilson loop. This was done by performing a supergauge transformation for an effectively unregularized original  
two-point function since we ignore order-$\varepsilon$ effects stemming from the field equations of motion. The transformed propagator possesses a
much softer ultraviolet behavior compared to the original one that we started from. It does even not require a regularization in order to perform 
integrations over the loop's contours and yields the expected superconformal $R$-invariant, according to superamplitude/super Wilson loop duality.
One of the further problems to study is to push the program beyond one-loop level and understand the supergauge transformation for two-loop
super Wilson loop. This would be much facilitated by an action functional for the chiral superconnection away from Wess-Zumino gauge,
whose existence is highly suggested by our results.

\vspace{5mm}

The work of AB was supported by the U.S. National Science Foundation under grants PHY-0757394  and PHY-1068286. AB would like
to thank J. Maldacena for the hospitality extended to him at the Institute for Advanced Study when this project was initiated. SCH gratefully 
acknowledges support from the Marvin L.~Goldberger Membership and from the National Science Foundation under grant PHY-0969448. 
We also would like to thank ECT* (Trento) and BIRS (Banff) for creating a fruitful atmosphere.


\end{document}